# On scaling laws at the phase transition of systems with divergent order parameter and/or internal length : the example of DNA denaturation

Sahin BUYUKDAGLI and Marc JOYEUX[a]

*Laboratoire de Spectrométrie Physique (CNRS UMR 5588),*
*Université Joseph Fourier, BP 87, 38402 St Martin d'Hères, France*

**Abstract** :   We used the Transfer-Integral method to compute, with an uncertainty smaller than 5%, the six fundamental characteristic exponents of two dynamical models for DNA thermal denaturation and investigate the validity of the scaling laws. Doubts concerning this point arise because the investigated systems (i) have a divergent internal length, (ii) are described by a divergent order parameter, (iii) are of dimension 1. We found that the assumption that the free energy can be described by a single homogeneous function is robust, despite the divergence of the order parameter, so that Rushbrooke's and Widom's identities are valid relations. Josephson's identity is instead not satisfied. This is probably due to the divergence of the internal length, which invalidates the assumption that the correlation length is solely responsible for singular contributions to thermodynamic quantities. Fisher's identity is even wronger. We showed that this is due to the $d=1$ dimensionality and obtained an alternative law, which is well satisfied at DNA thermal denaturation.

[a] email : Marc.JOYEUX@ujf-grenoble.fr

# I - Introduction

It has long been recognized that there are marked similarities between the phase transitions of very different systems : antiferromagnets, liquids, superconductors and ferroelectrics, to quote some of them, indeed all display a rather simple behavior in the region close to the critical point. A partial explanation comes from Landau's theory [1] and equivalent ones, like Van der Waals' equation for liquids, Weiss' molecular field theory for ferromagnets, Ornstein-Zernike equations, random phase approximations [2], and Ginzburg-Landau's equations for superconductors [3]. By supposing that the transition can be described by a so-called *order parameter* [2] and that the free energy can be expanded in power series in this parameter and the temperature gap $T_c - T$ (where $T_c$ is the critical temperature), these theories predict that most quantities (like the specific heat, the order parameter, the isothermal susceptibility, the correlation length and the correlation functions) display power laws in the neighborhood of the phase transition. Experiments done on many systems confirm the power laws predicted by Landau, but show that real critical exponents differ markedly from those predicted by the theory [4]. These experiments furthermore suggest that the various critical exponents are not independent but obey instead certain constraints. Phenomenological scenarii, which explain these observations, were proposed by Widom [5,6], Fisher [7-9], Kadanoff [4,10] and Domb and Hunter [11]. Based on the assumption that the free energy and/or the correlation length are homogeneous functions, these theories lead to the conclusion that all critical exponents can be expressed in terms of only two of them, thanks to so-called *scaling laws*. Later, a method known as the *Renormalization Group* theory, which is based on Wilson's idea that the critical point can be mapped onto a fixed point of a suitably chosen transformation of the system's Hamiltonian [12,13], has provided a conceptual framework for understanding scaling.

Yet, as far as we know, all the systems for which the validity of the scaling laws has been checked have two properties in common : (i) their phase transition is describable by a *finite* order parameter, and (ii) these systems *do not dissociate* at the critical temperature. The fact that the order parameter remains finite is essential for most theories, which assume that the free energy can be expanded in power series with respect to the order parameter and the temperature gap $T_c - T$. Obviously, this assumption no longer holds when the order parameter diverges at the critical point. Another central assumption of scaling theories is that the correlation length is solely responsible for singular contributions to extensive thermodynamic quantities. While this is certainly a reasonable assumption for bound systems, this might not be the case for dissociating ones. Indeed, a system that dissociates at the critical temperature possesses at least one physical internal length which increases infinitely at the critical point and might therefore contribute significantly to extensive thermodynamic quantities.

Whether the scaling laws are valid or not for systems with divergent order parameter and/or internal length is therefore an open question. The purpose of this paper is to address this question through the calculation of the characteristic exponents of two realistic dynamical models for DNA denaturation. This phase transition, which takes place when DNA solutions are heated, corresponds to the separation of the two DNA strands, that is, to the dissociation of the entangled polymers. Moreover, if the external stress depends explicitly on the distance between paired bases, then the corresponding order parameter diverges at the critical point. DNA denaturation models are therefore particularly well suited to investigate the applicability of the scaling laws to such unusual systems.

The remainder of this paper is organized as follows. The two dynamical models for DNA denaturation are briefly described in section II. The technique we used to compute the characteristic exponents is the Transfer-Integral (TI) method. The details of the calculations are sketched in section III. Finally, the applicability of the scaling laws to systems with divergent order parameter and/or internal length is discussed in section IV on the basis of the critical exponents that were obtained for the two models.

**II - The two dynamical models for DNA denaturation**

The potential energy $E_{pot}$ of the two dynamical models for DNA denaturation is of the general form

$$E_{pot} = \sum_k V(y_k) + W(y_k, y_{k+1}) + h f(y_k), \tag{II.1}$$

where $y_k$ denotes the position of the particle at site $k$, $V(y_k)$ is the on-site potential, $W(y_k, y_{k+1})$ the nearest-neighbour coupling between two successive particles, and $h f(y_k)$ plays the role of an externally applied constraint. The order parameter $m$ is obtained as the first derivative of the free energy with respect to the external field, that is, here

$$m = \frac{\partial}{\partial h}\left\{-k_B T \ln\left(\int \exp\left[-\frac{E_{pot}}{k_B T}\right] dy\right)\right\} = \frac{\int f(y) \exp\left[-\frac{E_{pot}}{k_B T}\right] dy}{\int \exp\left[-\frac{E_{pot}}{k_B T}\right] dy} = \langle f(y) \rangle. \tag{II.2}$$

In this work, we used $f(y_k) = y_k$ and $f(y_k) = y_k^2$, which lead to order parameters $m = \langle y \rangle$ and $m = \langle y^2 \rangle$, respectively.

The first model for DNA denaturation was proposed by Dauxois, Peyrard and Bishop (DPB) [14-17]. Expressions for the on-site potential and nearest-neighbor coupling are

$$\begin{aligned} V(y_k) &= D(1 - \exp[-a y_k])^2 \\ W(y_k, y_{k+1}) &= \frac{K}{2}(y_{k+1} - y_k)^2 (1 + \rho \exp[-\alpha(y_k + y_{k+1})]), \end{aligned} \tag{II.3}$$

where $y_k$ represents the transverse stretching of the hydrogen bond connecting the $k$th pair of bases. Numerical values of the coefficients are taken from Ref. [16], that is, $D$=0.03 eV, $a$=4.5 Å$^{-1}$, $K$=0.06 eV Å$^{-2}$, $\alpha$=0.35 Å$^{-1}$ and $\rho$=1. Thanks to the non-linear stacking interaction ($\rho > 0$), this model displays a much sharper transition at denaturation and is thus in better agreement with experiment than the older models on which it is based [18-20].

The second model for DNA denaturation was proposed by ourselves (JB) [21,22] to take into account the fact that stacking interactions are necessarily finite. For homogeneous sequences, it is of the form

$$V(y_k) = D\left(1 - \exp[-a\, y_k]\right)^2$$
$$W(y_k, y_{k+1}) = \frac{\Delta H}{2}\left(1 - \exp[-b(y_{k+1} - y_k)^2]\right) + K_b(y_{k+1} - y_k)^2 \,, \quad \text{(II.4)}$$

where $D$=0.04 eV, $a$=4.45 Å$^{-1}$, $\Delta H$=0.44 eV, $b$=0.10 Å$^{-2}$ and $K_b$=10$^{-5}$ eV Å$^{-2}$. The first term of $W(y_k, y_{k+1})$ describes the finite stacking interaction, while the second one models the stiffness of the sugar/phosphate backbone. Most interestingly, we were able, by introducing in this model the site-specific stacking enthalpies $\Delta H$ deduced from thermodynamic calculations [23], to reproduce the multi-step denaturation process that is experimentally observed for inhomogeneous DNA sequences.

**III - Tranfer-integral (TI) calculations**

The transfer-integral (TI) method (see for example Ref. [24] for a general description and Ref. [25] for a discussion regarding the applicability of the method to systems with unbound on-site potentials) consists in finding the eigenvalues $\lambda_k$ and eigenvectors $\phi_k$ of the symmetric TI operator, which satisfy

$$\int \phi_k(x) \exp\left[-\frac{V(x)+V(y)+2W(x,y)+h\,f(x)+h\,f(y)}{2k_BT}\right]dx = \lambda_k \phi_k(y). \tag{III.1}$$

For this purpose, we used the procedure described in Appendix B of Ref. [24], which is based on the diagonalization of a symmetric matrix with elements

$$M_{ij} = \delta_i^{1/2}\delta_j^{1/2} \exp\left[-\frac{V(u_i)+V(u_j)+2W(u_i,u_j)+h\,f(u_i)+h\,f(u_j)}{2k_BT}\right], \tag{III.2}$$

where the $u_i$ define a grid of non-necessarily equally-spaced values of the position coordinate and the $\delta_i$ stand for the intervals $\delta_i = (u_{i+1} - u_{i-1})/2$. The eigenvalues $\lambda_k$ of the symmetric TI operator coincide with the eigenvalues of the $\{M_{ij}\}$ matrix, while the eigenvectors $\phi_k$ of the symmetric TI operator are connected to the normalized eigenvectors $\{V_{k,i}\}$ of the $\{M_{ij}\}$ matrix through the relation $\phi_k(u_i) = \delta_i^{-1/2} V_{k,i}$. It is convenient to rewrite the eigenvalues in the form $\lambda_k = \exp[-\varepsilon_k/(k_BT)]$, and to label with a zero the quantities related to the largest eigenvalue (e.g. $\lambda_0$, $\varepsilon_0$, $\phi_0$ and $\{V_{0,i}\}$) and with a 1 those related to the second largest eigenvalue (e.g. $\lambda_1$, $\varepsilon_1$,...). In the thermodynamic limit of an infinite number of sites, the singular part of the specific heat $c_V$, the longitudinal correlation length $\xi$, the average $\langle g(y) \rangle$ of any function $g(y)$, and the static structure factor $S(q,T)$ are obtained according to

$$\begin{aligned}
c_V &= -T\frac{\partial^2 \varepsilon_0}{\partial T^2} \\
\xi &= \frac{l\,k_BT}{\varepsilon_1 - \varepsilon_0} \\
\langle g(y) \rangle &= \int g(y)|\phi_0(y)|^2 dy = \sum_i g(u_i) V_{0,i}^2 \\
S(q,T) &= \sum_{k\neq 0} R_k^2 \frac{\lambda_0^2 - \lambda_k^2}{\lambda_0^2 + \lambda_k^2 - 2\lambda_0\lambda_k \cos(q\,l)},
\end{aligned} \tag{III.3}$$

where *l* denotes a characteristic length of the system (we assumed without loss of generality that *l*=1 Å), and $R_k$ stands for the integral

$$R_k = \int f(y)\phi_k(y)\phi_0(y)dy = \sum_i f(u_i)V_{k,i}V_{0,i} \tag{III.4}$$

Note, that the derivative in the expression for $c_V$, as well as the derivative $dm/dh$ (see section IV), were computed from finite differences rather than from the complex expressions in Appendix B of Ref. [24].

The characteristic exponents were estimated by drawing log-log plots of the various quantities in Eq. (III.3) and measuring the slopes in the regions where power laws are satisfied. For obvious physical reasons, these regions do not extend far from the critical point. Unfortunately, numerical considerations also forbid the observation of these regions too close from the critical point. Indeed, an infinite range of *y* values would be needed to numerically converge the quantities in Eq. (III.3) at the critical point. Since the dimension of the $\{M_{ij}\}$ matrix is necessarily finite, numerical results can be accurate only up to a certain distance from it. Consequently, large grids of points extending to large values of *y* are mandatory for the interval on which power laws are observed to be broad enough to allow a precise estimation of the characteristic exponents. This point is absolutely crucial. For example, some of the characteristic exponents for the DPB model have already been reported [16]. However, the authors note that several quantities "diverge smoothly" at the transition, because of "transients which mask the leading-order asymptotics". As a consequence, they only provide rough estimates for the exponents, which sometimes differ by a factor 2 from exact values. In the light of our calculations, it appears that the so-called transients actually result from the numerical limitation mentioned above. In order to achieve better precision, we used grids of 4200 $u_i$ values regularly spaced between $y = -200/a$ and $y = 4000/a$ or, alternately, grids of the same length but with spacings which increase exponentially from $\delta_i = 0.2/a$ at $y \leq 0$ to $\delta_i = 4/a$ at $y = 5067/a$ (both grids lead essentially to the same result). We estimate on the basis of all our trials, that we were able to compute the exponents (see Eq. (IV.1) below) with an uncertainty smaller than 5 %.

**IV - Results and discussion**

A – Characteristic exponents

The six fundamental characteristic exponents α, β, γ, δ, η and ν (we omit the prime symbols although $T < T_c$) are traditionally defined according to

$$c_V \sim (T_c - T)^{-\alpha}$$
$$m \sim (T_c - T)^{\beta}$$
$$\frac{dm}{dh} \sim (T_c - T)^{-\gamma}$$
$$m \sim h^{1/\delta}$$
$$\xi \sim (T_c - T)^{-\nu}$$
$$S(q, T_c) \sim |q|^{\eta - 2}.$$

(IV.1)

α, β, γ and ν are computed at zero field (*h*=0), while δ and η are computed at critical temperature $T_c$. From the numerical point of view, $T_c$ was obtained as the temperature where the longitudinal correlation length ξ is maximum (at $h = 0$). With the exponentially spaced grid of length 4200, we calculated $T_c = 280.2934$ K for the DPB model and $T_c = 368.15$ K for the JB model. As indicated in Sect. III, the characteristic exponents were estimated by drawing log-log plots of the quantities in Eq. (IV.1) and measuring the slopes in the regions where power laws hold. For the sake of illustration, some plots for α, β, γ and ν are shown in Figs. 1 and 2. Fig. 1 deals with the DPB model with external constraint $f(y_k) = y_k$, while Fig. 2 deals with the JB model with the same external constraint. Measurement of the last two exponents δ and η were performed on similar plots, but with field ($h$) or wave-vector ($q$) abscissa. Note that we used two different external constraints for each model, namely $f(y_k) = y_k$ and $f(y_k) = y_k^2$, which correspond to order parameters $m = \langle y \rangle$ and $m = \langle y^2 \rangle$, respectively. Exponents β, γ and δ depend on the choice of the external constraint, while α, ν and η do not. The two sets of characteristics exponents that were obtained for each model are summarized in Table I.

At that point, two comments are in order. First, the characteristic exponent for specific heat, α, is significantly larger than 1 for both the DPB and the JB models. This confirms that both models predict a first-order phase transition at DNA denaturation temperature [16,21,26]. Moreover, the signs in Eq. (IV.1) were chosen such that that exponents are usually positive (although α and η are sometimes slightly negative). For the DPB and JB models, the order parameter $m$ however diverges at the critical point, so that β and δ are strongly negative.

B – <u>Rushbrooke's and Widom's identities</u>

The first two scaling laws, known as Rushbrooke's and Widom's identities, can be written in the form

$$\alpha + 2\beta + \gamma = 2$$
$$\gamma - \beta(\delta - 1) = 0,$$
(IV.2)

respectively. To obtain these relations, one just needs to assume that the singular part of the free energy, $f_{\text{sing}}$, can be described by a single homogenous function in $T_c - T$ and $h$, that is,

$$f_{\text{sing}}(T,h) = (T_c - T)^{2-\alpha} \, G\!\left(\frac{h}{(T_c - T)^{\Delta}}\right).$$
(IV.3)

Eq. (IV.2), as well as the additional relation $\Delta = \beta\delta$, then arise naturally from the interconnections between $f_{\text{sing}}$, $c_V$, $m$ and $dm/dh$ via thermodynamic derivatives. Eq. (IV.3) is actually a generalization of what is observed within the saddle-node approximation of the Ginzburg-Landau model, which leads to $f_{\text{sing}}(T,h) = (T_c - T)^2 \, G\!\left(h/(T_c - T)^{3/2}\right)$. The models investigated in this paper differ markedly from the Ginzburg-Landau one, but we checked that the *homogeneity assumption* of Eq. (IV.3) is nevertheless well satisfied. This is illustrated in Fig. 3, which shows the plots of $f_{\text{sing}}/(T_c - T)^{2-\alpha}$ versus the logarithm of $h/(T_c - T)^{\beta\delta}$ for the JB model with external constraint $f(y_k) = y_k$ and three values of $h$ ranging from $10^{-4} D$ to $10^{-6} D$. Note that in the TI formalism, $f_{\text{sing}}$ is obtained from

$$f_{\text{sing}}(T,h) = \varepsilon_0(T,h) - \varepsilon_0(T = T_c, h = 0). \qquad (\text{IV.4})$$

The fact that the points corresponding to different values of $h$ all lie on the same line indicates that the homogeneity assumption is correctly satisfied. It therefore comes as no surprise that Rushbrooke's and Widom's identities are also satisfied by the measured exponents. This is clearly seen in Table 2, which displays, for each polynome $\alpha + 2\beta + \gamma$ and $\gamma - \beta(\delta - 1)$, the value predicted by the corresponding scaling law (column 2) and those obtained from the measured values of the characteristic exponents (columns 3-6). Table 2 also provides qualitative uncertainties obtained by assuming that all exponents have additive 5% errors. It is seen that in all cases the values predicted by Rushbrooke's and Widom's identities lie well inside the uncertainty range.

### C – Josephson's identity

Josephson inequality [27,28] states that

$$\alpha + \nu d \geq 2 \;, \tag{IV.5}$$

where $d$ is the dimensionality of the system (here $d = 1$). This inequality converts to the equality known as Josephson's identity

$$\alpha + \nu d = 2 \;, \tag{IV.6}$$

if the *generalized homogeneity assumption* holds, that is, if (i) the only important length near the critical point is the correlation length $\xi$, and (ii) $\xi$ is solely responsible for all singular contributions to thermodynamic quantities. Note that if the *generalized homogeneity assumption* is satisfied, then the *homogeneity assumption* of Eq. (IV.3) is also satisfied, so that Rushbrooke's and Widom's identities are true.

Quite interestingly, examination of Table 2 shows that the computed exponents satisfy the inequality of Eq. (IV.5) but not Josephson's identity. Indeed, the difference between the computed values of $\alpha + \nu d$ and that predicted by the scaling law (*i.e.* 2) is larger than three times the 5% uncertainty for both models. This indicates that, in contrast with many systems, the generalized homogeneity assumption does not hold for DNA denaturation. As we anticipated in the Introduction, this is not unexpected for systems which dissociate at the critical point. Indeed, these systems possess at least one physical internal length which increases infinitely at the critical temperature, so that it is no longer justified to assume that everything is a function only of the ratio of a typical finite microscopic length to the correlation length $\xi$. We unsuccessfully tried to figure out, on the basis of the numerical values reported in Tables 1, what quantity could replace the correlation length $\xi$ in the generalized homogeneity assumption (this quantity should obviously have length dimension and a characteristic exponent equal to $2 - \alpha$).

D – <u>Fisher's identity</u>

Fisher's identity connects γ, η and ν according to

$$\gamma - \nu(2 - \eta) = 0 \qquad (IV.7)$$

Examination of Table 2 shows that this equality is very far from being satisfied by the models for DNA denaturation. The reason for these discrepancies is that Fisher's identity is based on the assumption that the correlation function

$$G(x) = \langle f(y_j) f(y_{j+x}) \rangle - \langle f(y) \rangle^2 \qquad (IV.8)$$

falls off, close to the critical temperature, as

$$G(x) \sim \frac{1}{x^{d-2+\eta}} \ . \qquad (IV.9)$$

While correct for the Ginzburg-Landau Hamiltonian with $d \geq 2$, this assumption is just wrong for a system with $d = 1$ and $\eta \approx 0$ because, for these values of $d$ and η, Eq. (IV.9) diverges with increasing values of $x$. In the TI formalism, $G(x)$ may be obtained from [24]

$$\langle f(y_j) f(y_{j+x}) \rangle = \sum_k R_k^2 \left( \frac{\lambda_k}{\lambda_0} \right)^x \ . \qquad (IV.10)$$

Numerically, we found that Eq. (IV.10) actually leads to constant values of $G(x)$ for the two investigated models close to the critical temperature. Evaluating these constants at $x = 0$, one gets

$$G(x) = \langle f(y)^2 \rangle - \langle f(y) \rangle^2 \sim \langle f(y)^2 \rangle \ . \qquad (IV.11)$$

Writing, as usual, that

$$\frac{dm}{dh} \sim \int_0^\xi G(x)\,dx \ , \qquad (IV.12)$$

one thus obtains, instead of Fisher'identity, the relation

$$\gamma = \mu + \nu \ , \qquad (IV.13)$$

where μ is the characteristic exponent for $\langle f(y)^2 \rangle$

$$\left\langle f(y)^2 \right\rangle \sim (T_c - T)^{-\mu} \;. \tag{IV.14}$$

The measured values of μ are reported in Table 1 (note that the exponent μ for $f(y_k) = y_k$ is the opposite of β for $f(y_k) = y_k^2$). The validity of the scaling rule in Eq. (IV.13) is checked in Table 2. The agreement is excellent.

**V - Conclusion**

We investigated the validity of the scaling rules for two dynamical models of DNA thermal denaturation. These models indeed display several characteristics, which shed doubts on this question : (i) the distance between paired bases, that is, the physical length in terms of which the Hamiltonian is expressed, diverges at the melting temperature, (ii) the expressions we assumed for the external constraint lead to order parameters, which also diverge at the critical temperature, (iii) the dimensionality is $d = 1$. Conclusions are :

- the assumption that the free energy can be described by a single homogeneous function seems to be rather robust, despite the divergence of the order parameter. Consequently, Rushbrooke's and Widom's identities are valid relations.

- Josephson's identity is instead not satisfied. We argued that this is probably due to the divergent internal length, which invalidates the assumption that the correlation length is solely responsible for singular contributions to thermodynamic quantities.

- Fisher's identity is still farther from being satisfied. We showed that this is due to the $d = 1$ dimensionality and obtained an alternative law, which is well satisfied at DNA thermal denaturation.

Of course, one cannot derive general conclusions from a single study, and additional work is certainly needed to ascertain the robustness of the homogeneity assumption for free energy and/or improve Josephson's identity. This work still indicates that scaling laws must be handle with care when dealing with systems with unusual characteristics.

# TABLE CAPTIONS

**Table 1** : Values of the six fundamental characteristic exponents α, β, γ, δ, η and ν for the DPB and JB models with external constraints $f(y_k) = y_k$ and $f(y_k) = y_k^2$. The seventh exponent μ characterizes the behavior of $\langle f(y)^2 \rangle$ close to the critical temperature (see Sect. IV-D).

**Table 2** : Values of $\alpha + 2\beta + \gamma$, $\gamma - \beta(\delta - 1)$, $\alpha + \nu d$, $\gamma - \nu(2 - \eta)$ and $\gamma - (\mu + \nu)$ predicted by scaling laws (column 2) and obtained from the measured characteristic exponents reported in Table 1 for the DPB and JB models with external constraints $f(y_k) = y_k$ and $f(y_k) = y_k^2$ (columns 3-6). The uncertainties correspond to additive 5% errors for all the exponents. The last scaling law, $\gamma - (\mu + \nu) = 0$, is introduced in Sect. IV-D.

**FIGURE CAPTIONS**

**Figure 1** (color online) : Log-log plots used to determine the critical exponents α, β, γ and ν for the DPB model with external constraint $f(y_k) = y_k$.

**Figure 2** (color online) : Log-log plots used to determine the critical exponents α, β, γ and ν for the JB model with external constraint $f(y_k) = y_k$.

**Figure 3** (color online) : Plots of $f_{sing}/(T_c - T)^{2-\alpha}$ versus the logarithm of $h/(T_c - T)^{\beta\delta}$ for the JB model with external constraint $f(y_k) = y_k$ and three values of $h$ ranging from $10^{-4} D$ to $10^{-6} D$. $f_{sing}$ and $h$ are expressed in units of $D$. The fact that the points corresponding to different values of $h$ all lie on the same line indicates that the homogeneity assumption of Eq. (IV.3) is correctly satisfied by the model.

**TABLE 1**

|   | DPB model | | JB model | |
|---|---|---|---|---|
|   | $f(y) = y$ | $f(y) = y^2$ | $f(y) = y$ | $f(y) = y^2$ |
| α | 1.45 | 1.45 | 1.13 | 1.13 |
| β | -1.07 | -1.72 | -1.31 | -2.11 |
| γ | 2.86 | 4.00 | 3.33 | 4.82 |
| δ | -1.66 | -1.39 | -1.58 | -1.35 |
| η | 0.01 | 0.01 | 0.02 | 0.02 |
| ν | 1.12 | 1.12 | 1.23 | 1.23 |
| μ | 1.72 | 2.98 | 2.11 | 3.52 |

**TABLE 2**

|  | scaling law | DPB model | | JB model | |
|---|---|---|---|---|---|
|  |  | $f(y) = y$ | $f(y) = y^2$ | $f(y) = y$ | $f(y) = y^2$ |
| Rushbrooke : $\alpha + 2\beta + \gamma$ | 2 | 2.17±0.32 | 2.01±0.44 | 1.84±0.35 | 1.73±0.51 |
| Widom : $\gamma - \beta(\delta - 1)$ | 0 | 0.01±0.37 | -0.11±0.53 | -0.05±0.44 | -0.14±0.63 |
| Josephson : $\alpha + \nu d$ | 2 | 2.57±0.13 | 2.57±0.13 | 2.36±0.12 | 2.36±0.12 |
| Fisher : $\gamma - \nu(2 - \eta)$ | 0 | 0.63±0.25 | 1.77±0.31 | 0.89±0.29 | 2.38±0.36 |
| $\gamma - (\mu + \nu)$ | 0 | 0.02±0.28 | -0.10±0.40 | -0.01±0.33 | 0.07±0.48 |

**FIGURE 1**

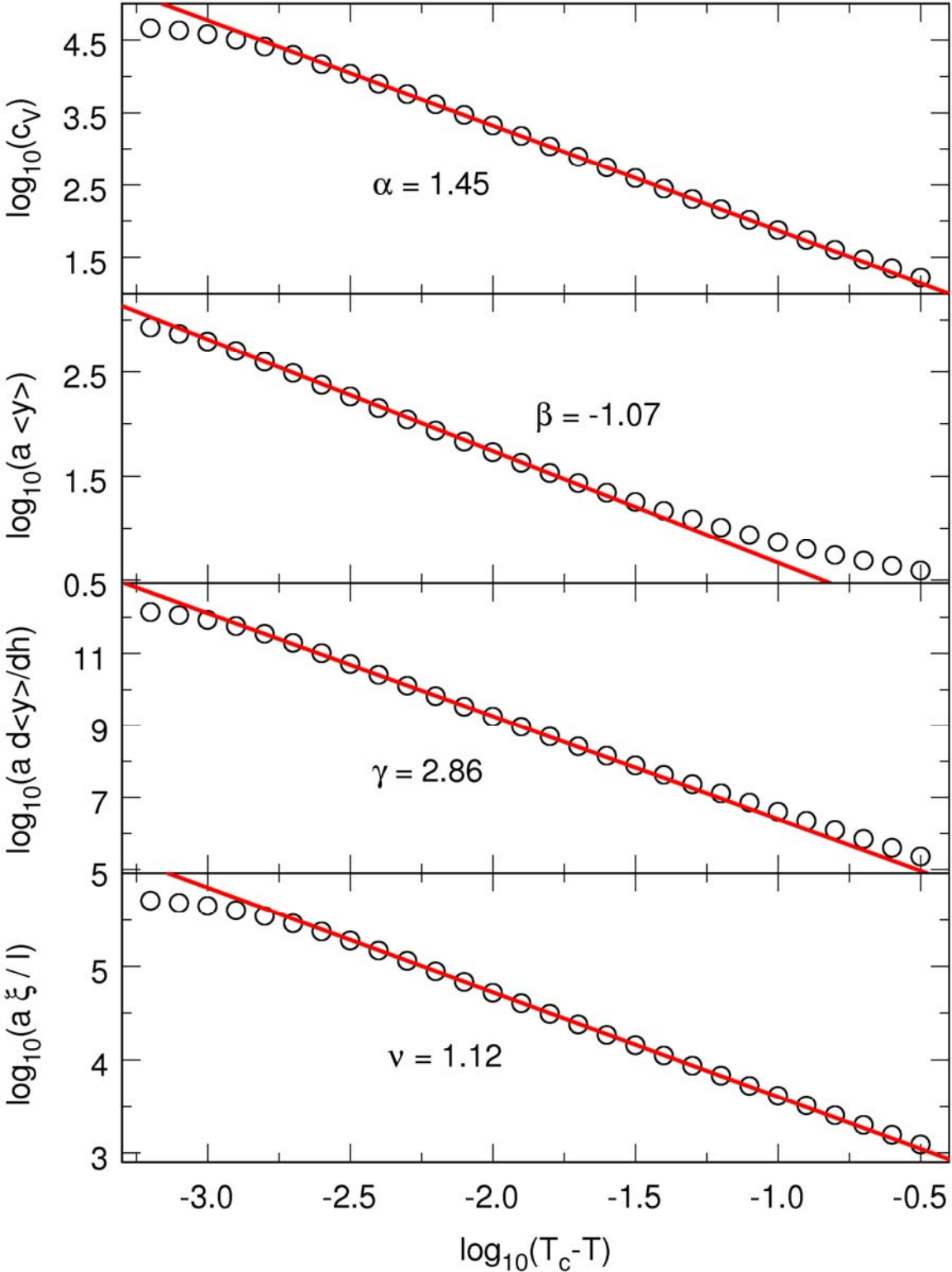

**FIGURE 2**

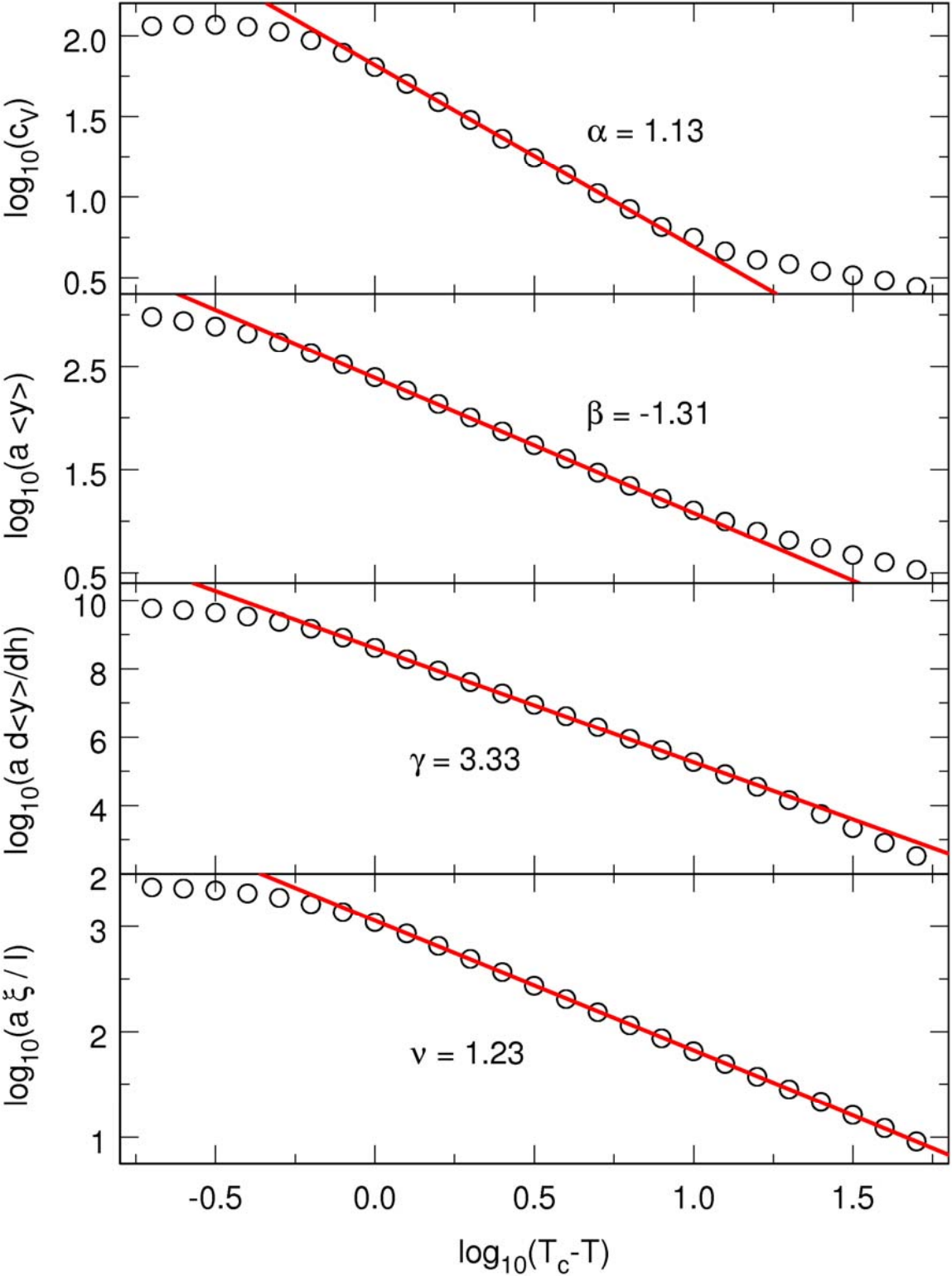

**FIGURE 3**

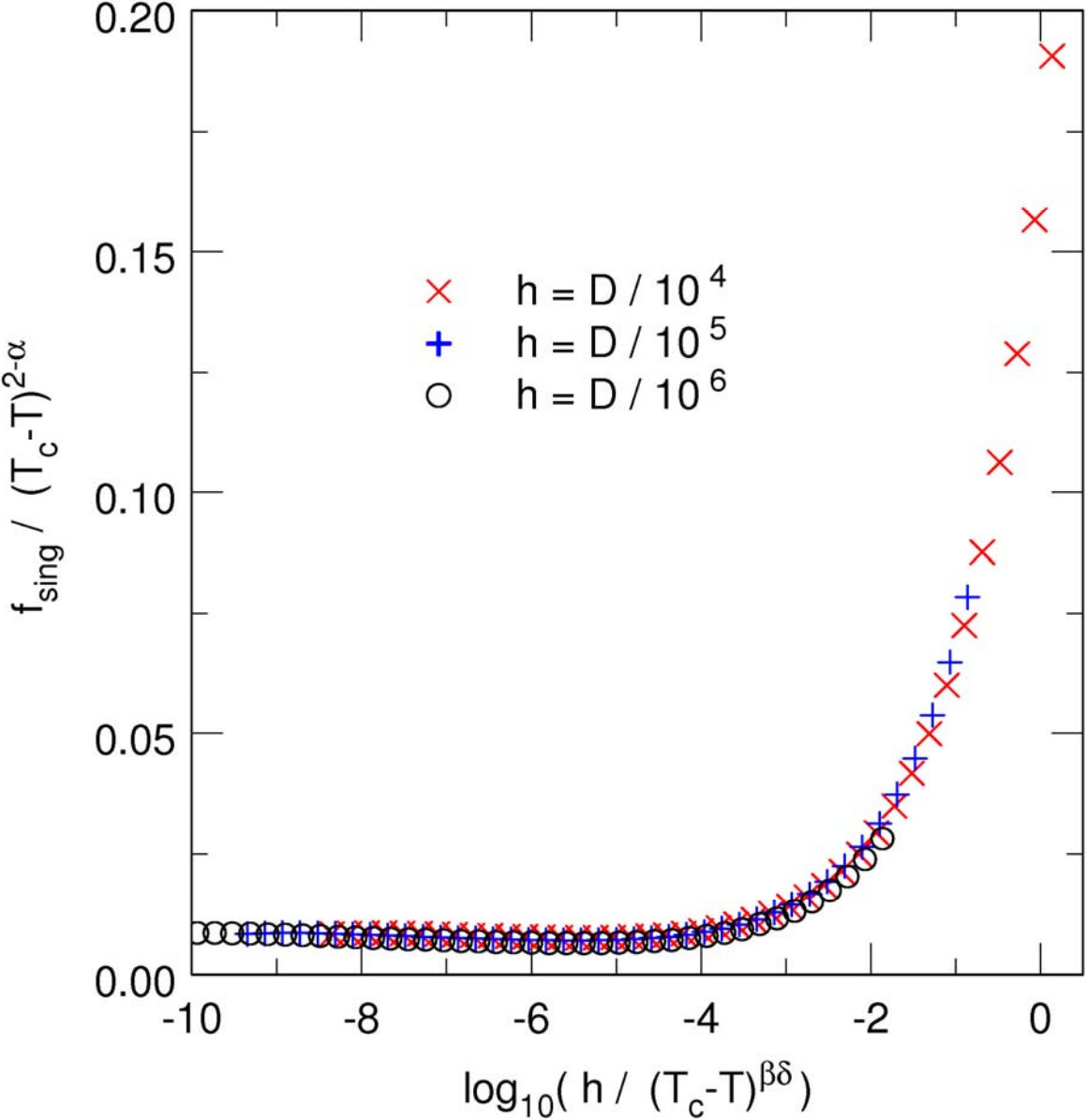